\documentclass[aps,prl,twocolumn,showpacs,amsmath,amssymb,floatfix]{revtex4-1}

\usepackage{float}
\usepackage{epsfig}
\usepackage{amsmath}
\usepackage{amssymb}
\usepackage{bm}
\usepackage{graphicx}
\usepackage{graphicx,rotating}
\usepackage{ulem}
\usepackage{epic}
\usepackage{color}
\usepackage{dashrule}
\begin{document}
\author{Jian-Xing Li}\email{Jian-Xing.Li@mpi-hd.mpg.de}
\affiliation{Max-Planck-Institut f\"{u}r Kernphysik, Saupfercheckweg 1,
69029 Heidelberg, Germany}
\author{Karen Z. Hatsagortsyan}\email{k.hatsagortsyan@mpi-hd.mpg.de}
\affiliation{Max-Planck-Institut f\"{u}r Kernphysik, Saupfercheckweg 1,
69029 Heidelberg, Germany}
\author{Christoph H. Keitel}\email{keitel@mpi-hd.mpg.de}
\affiliation{Max-Planck-Institut f\"{u}r Kernphysik, Saupfercheckweg 1,
69029 Heidelberg, Germany}


\title{Robust signatures of quantum radiation reaction in focused ultrashort laser pulses }

\begin{abstract}
Radiation reaction effects in the interaction of an electron bunch with a superstrong focused ultrashort laser pulse are investigated in the quantum radiation dominated regime. The angle-resolved Compton scattering spectra are calculated in laser pulses of variable duration using a semi-classical description for the radiation dominated dynamics and a full quantum treatment for the emitted radiation. In dependence of the laser pulse duration we find signatures of quantum radiation reaction in the radiation spectra, which are characteristic for the focused laser beam and visible in the qualitative behaviour of both the angular spread and the spectral bandwidth of the radiation spectra. The signatures are robust with respect to the variation of the electron and laser beam parameters in a large range. They fully differ qualitatively from those in the classical radiation reaction regime and are measurable with presently available laser technology.

\end{abstract}
\maketitle

Recent advances of strong field laser techniques have enabled the development of novel all-optical x-/$\gamma$-ray radiation sources \cite{Phuoc_2012,Corde_2013,RMP_2012}, being beneficial for broad applications, see, e.g., \cite{Achterhold_2013,ELI}. In particular, x-/$\gamma$-rays are achieved via Compton backscattering of laser radiation off a relativistic electron beam \cite{Chen_2013,Powers_2014}. In superstrong laser fields the Compton scattering acquires nonlinear characteristics due to multiple laser photon absorption \cite{Goldman_1964,Nikishov_1964,Brown_1964,Ritus_1985}. Moreover, in strong fields multiple photon emission during a laser period can be very probable. Consequently, the electron dynamics can be modified due to radiation, arising radiation reaction (RR) effects \cite{Koga_2005,RMP_2012}. Although the RR problem is discussed since the early days of classical \cite{Abraham_1905,Lorentz_1909,Dirac_1938,Landau_2} and quantum \cite{Heitler_1941,Wilson_1941,Heitler_1984,Jauch_1976} electrodynamics, the theory 
has not yet been tested experimentally. The development of the 
Extreme Light Infrastructure \cite{ELI} opens new perspectives to observe RR effects in laser-matter interaction at extreme conditions and revived the interest to this problem \cite{DiPiazza_2009,Sokolov_2009,DiPiazza_2010,
Sokolov_2010b,Sokolov_2010,Sokolov_2011,Bulanov_2011pre,
Harvey_2011a,Bulanov_2012,Pegoraro_2012, Harvey_2012,Thomas_2012,Schlegel_2012,Pandit_2012,Mackenroth_2013,Capdessus_2013,Kumar_2013,Tamburini_2014,Blackburn_2014,RMP_2012}.

The quantum effects in multiphoton Compton scattering (the photon recoil and spin effects) are determined by the invariant parameter $\chi\equiv |e|\sqrt{(F_{\mu\nu}p^{\nu})^2}/m^3$ \cite{Reiss_1962, Ritus_1985}, where $F_{\mu\nu}$ is the field tensor, $p^{\nu}=(\varepsilon,\textbf{p})$ the incoming electron 4-momentum, and $e$ and $m$ are the electron charge and mass, respectively (Planck units $\hbar=c=1$ are used throughout). In particular, the recoil of the emitted photon can be estimated as $\chi \sim \omega/\varepsilon$, with the emitted photon energy $\omega$ \cite{RMP_2012} and the quantum regime of radiation setting in at $\chi\gtrsim 1$. Physically, the parameter $\chi$ equals the ratio $\chi=E'/E_S$ of the laser field $E'$ in the electron rest frame over the critical Schwinger field $E_S=m^2/|e|$ \cite{Schwinger_1951}. The laser intensity corresponding to the Schwinger field is $I_S= E_S^2/(8 \pi)\approx 2.3\times 10^{29}$ W/cm$^2$ which cannot be reached with realistic lasers \cite{Yanovsky_2008,ELI}. 
However, a relativistic electron counterpropagating with the laser field may experience the Schwinger field in its rest-frame $E' \approx 2 \gamma E$, i.e., due to the Lorentz-boost of the laser field $E$ with the Lorentz $\gamma$-factor. In this case $\chi \approx 2\gamma \xi \omega_0/m$, with the laser frequency $\omega_0$ and the invariant laser parameter $\xi\equiv |e|E/(m\omega_0)$.

As the probability of emitting a photon in a so-called formation length is of the order of the fine structure constant $\alpha$ \cite{Bolotovskii_1986} and since one laser period contains about $\xi$ formation lengths \cite{Ritus_1985}, the average number of photons emitted by an electron in a laser period is $N_{ph} \sim \alpha \xi$, and the electron energy loss due to radiation yields $\Delta \varepsilon \sim \alpha \xi\chi \varepsilon$. Thus, the radiation-dominated regime (RDR) can be characterized by the parameter $R \equiv\alpha \xi \chi \gtrsim 1$ \cite{RMP_2012}. For available petawatt infrared lasers, $I\lesssim 3\times10^{22}$ W/cm$^2$ \cite{Yanovsky_2008} ($\xi \lesssim 200$), the RDR is achievable only if $\chi\gtrsim 1$, i.e., in the quantum regime of interaction.
 A peculiar RDR has been identified in \cite{DiPiazza_2009}, when RR in the classical regime becomes prominent at  $\xi\sim 100$, though in a rather narrow range of parameters near the, so-called, reflection condition $\xi\approx 2\gamma$. One concluded  in \cite{Thomas_2012} that the RR effects in the classical regime are mostly detectable through measurements of electron beam properties. Various modifications of the radiation spectrum in the quantum RDR of Compton scattering were put forward in \cite{Sokolov_2010b,DiPiazza_2010} which, however, are not easily discernible in an experiment and require an accurate quantitative measurement. The role of stochastic effects in the quantum RDR was further studied. Those yield an increase of the electron energy and transverse spreading \cite{Neitz_2013,Green_2014} as well as an increased output of high energy photons \cite{DiPiazza_2010,Blackburn_2014}.

The aim of this work is to identify signatures of RR for Compton radiation spectra in the quantum RDR, which can be easily detectable in an experiment due to distinct qualitative characteristics. The parameters $R \gtrsim 1$ and $\chi \lesssim 1$ are employed to ensure that pair production effects are negligible while quantum recoil effects remain important. We investigate features of the angle-resolved spectra of Compton radiation when an ultrarelativistic electron beam counterpropagates with a strong focused ultrashort laser pulse of variable duration. In particular, with increasing laser pulse duration the angular spread of the main photon emission region (MPER) is shown to initially rise in a narrow range due to laser focusing and then continuously decreases because of quantum RR. This unique behaviour does not exist in the classical RR regime. The spectral bandwidths of the radiation in the quantum and classical regimes both monotonously decrease when the laser pulse duration is increased, but the former is by orders of magnitude larger due to much stronger RR effects. The mentioned qualitative behaviours are observed in a broad range of laser and electron parameters. The electron dynamics including RR is described by classical equations of motion \cite{Sokolov_2010b}, while the emitted radiation is calculated quantum mechanically \cite{Ritus_1985}. The simple quasiclassical approach for the electron dynamics is justified as the electron's de-Broglie wavelength is much smaller than the laser wavelength, and it allows to explore the role of the laser focusing effect in the quantum RDR.

Usually, high laser intensities are obtained by focusing a laser beam to a spot size of the order of the wavelength. When additionally the laser pulse is of duration of only few cycles, then the well known paraxial approximation 
\cite{Salamin_2006,Salamin_2002,Salamin_2007,Li_2009a,Li_2009b} is not suitable for its description. In this case
the small diffraction parameter $(k_0w_0)^{-1}$ is of the same order of magnitude as the temporal parameter $(\omega_0 \tau_0)^{-1}$, where $k_0$, $w_0$ and $\tau_0$ are the wave vector, the waist radius and the pulse duration of the laser beam, respectively, and the approximate solution of Maxwell equations should treat both parameters on equal footing. We consider a  circularly polarized (CP) focused ultrashort laser pulse propagating along the $z$-direction. The field of the laser pulse is derived in Supplemental Material \cite{Suppl_material} analogous to \cite{Esarey_1995}.   
Note that the temporal envelope of the laser beam is not factorized in this solution.

We describe RR as emission of multiple photons during the electron motion in a laser field when the electron dynamics is accordingly modified following the photon emissions. In superstrong laser fields $\xi\gg 1$, the coherence length of the photon emission is much smaller than the laser wavelength \cite{Ritus_1985} and the photon emission probability is determined by the local electron trajectory, consequently, by the local value of the parameter $\chi$.  The differential probability per unit phase interval is \cite{Ritus_1985,Sokolov_2010}:
\begin{eqnarray}
\frac{d W_{fi}}{d \eta d\tilde{\omega}}=\frac{\alpha \tilde{\chi} m^2[\int_{\tilde{\omega}_{r}}^{\infty} K_{5/3}(x)dx+\tilde{\omega}  \tilde{\omega}_{r} \tilde{\chi}^2 K_{2/3}(\omega_{r})]}{\sqrt{3}\pi(k_{0i}\cdot p_i)},
\label{W}
\end{eqnarray}
where $\eta = \omega_0 t -k_0 z$, $\tilde{\omega} =k_{0i}\cdot k_i/(\tilde{\chi}\,k_{0i}\cdot p_i)$ is the normalized emitted photon energy, $\tilde{\chi}=3\chi/2$, $k_{0i}$, $k_i$ and $p_i$ are the four-vectors of the driving laser photon, the emitted photon and the electron, respectively,  and $\tilde{\omega}_r  =\tilde{\omega}/\rho_0$ with recoil parameter $\rho_0 =1-\tilde{\chi}\tilde{\omega}$ (in the classical limit $\rho_0\approx 1$). The characteristic energy of the emitted photon is determined from the relation $\tilde{\omega}_r\sim 1$ and yields the cut-off frequency $\omega_c\sim \chi\varepsilon/(2/3+\chi)$. 
The rate of the electron radiation loss is ${\cal I}=\int  d\tilde{\omega}(k_{0i}\cdot k_i)d W_{fi}/(d \eta d\tilde{\omega})$.
Implementing the radiation losses due to quantum RR into the electron classical dynamics   leads to the following equation of motion \cite{Sokolov_2010}:
\begin{eqnarray}
\frac{d p^{\alpha}}{d\tau}=\frac{e}{m }F^{\alpha\beta}p_\beta-\frac{{\cal I}}{m }p^{\alpha}+\tau_c\frac{{\cal I}}{{\cal I}_c}F^{\alpha\beta}F_{ \beta\gamma}p^{\gamma},
\label{eq2}
\end{eqnarray}
where $\tau$ is the proper time, $\tau_c\equiv 2e^2/(3m)$ and ${\cal I}_{c}=2\alpha\omega^2\xi^2$ is the classical radiation loss rate.  

\begin{figure}
\includegraphics[width=8cm]{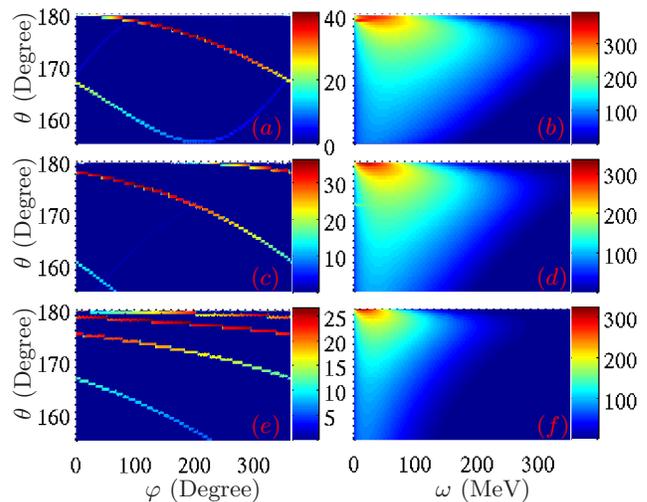}
\begin{picture}(0,0)(100,10)
\put(-50,139){\color{red}$(a)$}
\put(58,139){\color{red}$(b)$}
\put(-50,82){\color{red}$(c)$}
\put(58,82){\color{red}$(d)$}
\put(-50,26){\color{red}$(e)$}
\put(58,26){\color{red}$(f)$}

\put(-90,2){$\varphi$ (Degree)}
\put(20,2){$\omega$ (MeV)}

\put(-140,143){\begin{sideways}$\theta$ (Degree)\end{sideways}}
\put(-140,88){\begin{sideways}$\theta$ (Degree)\end{sideways}}
\put(-140,31){\begin{sideways}$\theta$ (Degree)\end{sideways}}

\end{picture}

\caption{(Color online) The angle-resolved spectra of electron radiation in laser pulses of various durations: the left column displays d$\varepsilon$/d$\Omega$ [GeV/sr] and the right d$\varepsilon$/d$\omega$d$\Omega$ [1/sr] for (a)-(b) $\tau_0=T_0$, (c)-(d) $\tau_0=1.5T_0$, and (e)-(f) $\tau_0=5T_0$. The laser wavelength is $\lambda_0 = 1\mu$m while $w_0$ = 10$\lambda_0$, $\phi=0$, $\xi$=230, and $\gamma_0$=1000.}
\label{schem1}
\end{figure}

To study signatures of quantum RDR, we employ electrons with an initial energy of 500 MeV to interact with counterpropagating strong laser pulses of peak intensity $I\approx 7\times 10^{22}$ W/cm$^2$ ($\xi=230$, $\chi \approx 0.6$ and $R\approx 1$) with various durations. 
We firstly study the radiation spectra of a single electron and then proceed with the case of an electron beam.
The radiation spectra in laser pulses of $\tau_0=T_0, 1.5T_0, 5T_0$, with the laser period $T_0$, are illustrated in Fig.~\ref{schem1}. The photon emission direction is determined by the polar angle $\theta$ with respect to the laser propagation direction and the azimuthal angle $\varphi$ with respect to the $x-z$ polarization plane.
The distribution within $155^{\circ}\leq\theta\leq180^{\circ}$ is investigated, where the emission is mostly concentrated. The left column shows the angular distribution of the radiation energy,
d$\varepsilon$/d$\Omega$, with the solid angle $\Omega$. The polar angle spread for MPER (roughly the red color part) is the largest for $\tau_0=1.5T_0$, increasing when changing $\tau_0=T_0$ to $\tau_0=1.5T_0$, and decreasing with further rising pulse duration. For each $\theta$, there is a relevant $\varphi=\varphi_m$ where d$\varepsilon$/d$\Omega$ is maximal. The corresponding radiation spectrum, $d^2\varepsilon$/(d$\omega$d$\Omega)$ at $\varphi=\varphi_m$ is shown in the right column. While the value of $\varphi_m$ depends on the laser carrier-envelope-phase (CEP), the spectral intensity at this phase does not.

\begin{figure}
\includegraphics[width=8cm]{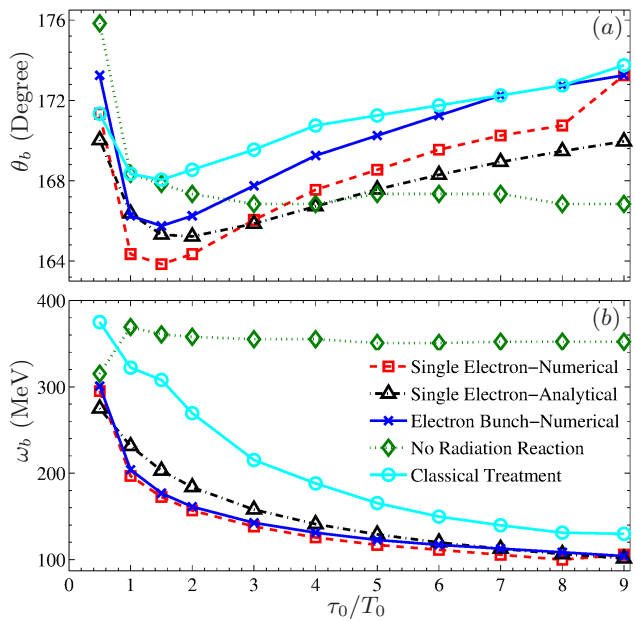}
\begin{picture}(0,0)(100,10)
\put(80,222){$(a)$}
\put(80,112){$(b)$}

\put(-20,3){$\tau_0/T_0$}
\put(-140,170){\begin{sideways}$\theta_b$ (Degree)\end{sideways}}
\put(-140,60){\begin{sideways}$\omega_b$ (MeV)\end{sideways}}
\end{picture}
\caption{(Color online). The quantum RR signatures in the quantum RDR. The boundary angle $\theta_b$ (a) and the boundary frequency $\omega_b$ (b) of the emitted photons are displayed in dependence on the laser pulse duration. The parameters are equal to those of Fig.~\ref{schem1}.}
\label{schem2}
\end{figure}

\begin{figure}

\includegraphics[width=8cm]{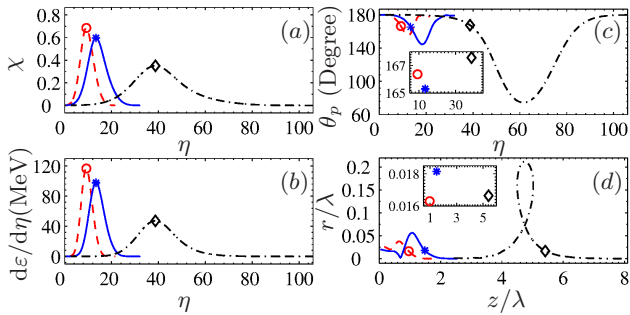}
\begin{picture}(0,0)(100,10)
\put(-36.5,105){$(a)$}
\put(82,105){$(c)$}
\put(-36.5,48){$(b)$}
\put(81.5,48){$(d)$}

\put(-75,3){$\eta$}
\put(42,3){$z/\lambda$}
\put(-75,62){$\eta$}
\put(44,62){$\eta$}

\put(-139,92){\begin{sideways}$\chi$\end{sideways}}
\put(-139,13){\begin{sideways}d$\varepsilon$/d$\eta$(MeV)\end{sideways}}
\put(-22,71){\begin{sideways}$\theta_p$ (Degree)\end{sideways}}
\put(-23,32){\begin{sideways}$r/\lambda$\end{sideways}}

\end{picture}
\caption{(Color online). The electron dynamics in counterpropagating laser pulses of various durations. The red dashed, blue solid, and black dash-dotted curves correspond to the laser pulse durations $\tau_0=T_0$, $\tau_0=1.5T_0$, and $\tau_0=5T_0$, respectively. Other parameters are equal to those of Fig.~\ref{schem1}. The marks point out the places where the corresponding $\chi$ is maximal. The insets in (c) and (d) show the details of the main plot.}
\label{schem4}
\end{figure}

The dependences of both angular distribution and spectral bandwidth of MPER on the laser pulse duration are summarized in Fig.~\ref{schem2}. The MPER is defined via the polar angular spread, $\Delta \theta$ = $180^{\circ}-\theta_b$, with a boundary angle $\theta_b$, where d$\varepsilon$/d$\Omega |_{\varphi=\varphi_m, \theta=\theta_b}= (d\varepsilon$/d$\Omega |_{max})/2$. The corresponding spectral bandwidth $\Delta \omega$ of the radiation is defined as  $\Delta \omega$ = $\omega_b-0$, with a boundary frequency $\omega_b$, where $d^2\varepsilon$/(d$\omega$d$\Omega)|_{\theta=\theta_b, \omega=\omega_b}=(d^2\varepsilon$/(d$\omega$d$\Omega)|_{\theta = \theta_b, \max}$)/2.  
As illustrated in Fig.~\ref{schem2}(a), the boundary angle $\theta_b$ firstly decreases (i.e., $\Delta \theta$ increases) in an ultrashort pulse range $\tau_0\lesssim 1.5T_0$, no matter whether RR effects are included, 
because of the laser focusing effect (explained later in Fig.~\ref{schem4}). When the laser pulse duration is further increased, the boundary angle $\theta_b$ monotonously rises (i.e., $\Delta \theta$ decreases) if the quantum RR effect is included (red dashed curve with square marks), and it almost remains unaltered if the RR effect is artificially removed in Eq.~\ref{eq2} (green dotted curve with diamond marks). In Fig.~\ref{schem2}(b), the spectral bandwidth $\Delta \omega = \omega_b$ monotonously decreases with rising laser pulse duration if quantum RR is taken into account, and it is almost constant when RR is neglected. Therefore, the signatures of RR in the emission spectra are easily distinguishable. Note that if the quantum effects (quantum recoil effect and the comparatively negligible radiation effect induced by electron spin) are artificially removed, keeping however RR fully classically, the variation dynamics within the MPER remains qualitatively similar but with apparent quantitative differences (cyan solid curves with circle marks).

\begin{figure}
\includegraphics[width=8cm]{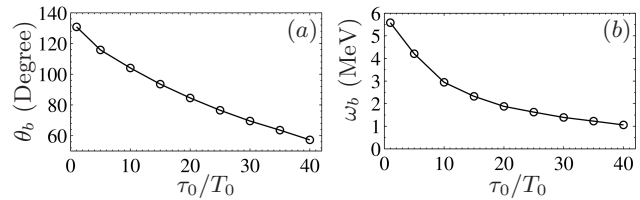}
\begin{picture}(0,0)(100,10)
\put(-78,3){$\tau_0/T_0$}
\put(40,3){$\tau_0/T_0$}
\put(-140,22){\begin{sideways}$\theta_b$ (Degree)\end{sideways}}
\put(-17,30){\begin{sideways}$\omega_b$ (MeV)\end{sideways}}
\put(-36.5,62){$(a)$}
\put(83,62){$(b)$}

\end{picture}

\caption{The RR signatures in the classical RR regime. The variation of (a) the boundary angle $\theta_b$ and (b) the boundary frequency $\omega_b$ is displayed versus the laser pulse duration. $\xi$ = 100, $\gamma$ = 100, and the other parameters are equal to those of Fig.~\ref{schem1}. }  
\label{schem3}
\end{figure}

Furthermore, our quantitative analysis shows that the discussed signatures of quantum RDR (when $R \gtrsim 1$ and $\chi \sim 1$) are clearly measurable in a broad range of parameters $\xi \lesssim \gamma \lesssim 20\xi$. In the extreme conditions $\gamma \gg \xi$ or $\gamma \ll \xi$, either the electron deflection angle with respect to the laser propagation axis $\theta_e \sim \xi/\gamma$ is vanishing and the photon emission is mostly 
along a polar angle $\theta = 180^{\circ}$, or the electron is quickly reflected by the laser pulse and emits within a very narrow angular spread near  $\theta = 0^{\circ}$. Thus, the signatures require $\xi\sim \gamma$, and in this case the RDR regime $R \gtrsim 1$ is equivalent to the quantum regime $\chi \gtrsim 1$. While the classical RR regime $\chi \ll 1$ is tantamount to the out-of-RDR limit $R\ll 1$. Let us briefly discuss how the RR signatures considered here behave in a typical classical RR regime. We employ 
$\gamma=100$ and $\xi=100$ to obtain $\chi\approx 10^{-2}$ and $R\approx 10^{-2}$. As presented in Fig.~\ref{schem3}, the behaviour of the boundary angle $\theta_b$ vs the pulse duration is qualitatively different from that in the quantum RDR discussed above, since the RR effect is much smaller in the classical regime. Although, the boundary frequency $\omega_b$ monotonously reduces with the increase of the laser pulse as well, the magnitude of the spectral bandwidth is roughly two orders of magnitude lower than that in the quantum RDR.

To explain the properties of the MPER highlighted in Fig.~\ref{schem2}, the corresponding electron dynamics is analysed in Fig.~\ref{schem4}. The maximal value of the parameter $\chi$ reduces with the increase of the pulse duration due to RR, as shown in Fig.~\ref{schem4}(a). The radiation loss rate, $d\varepsilon/d\eta$, follows the behaviour of the $\chi$-parameter as expected, cf. Figs.~\ref{schem4}(a) and (b), which also can be analytically estimated as follows. 
The radiation loss in the coherence length $\Delta \eta_{coh}\sim 2 \pi /\xi$ can be estimated as $\Delta \varepsilon\sim \alpha \omega_c$, with the emission cut-off frequency $\omega_c\sim m\gamma \chi/(2/3+\chi)$, which yields the  radiation loss rate ${\cal I}=$ d$\varepsilon$/d$\eta\sim \Delta \varepsilon/\Delta\eta_{coh} \sim\alpha \omega_c\xi/(2 \pi)\propto\chi^2$,
in line with the numerical result.
The instantaneous momentum directions of the scattered electron are presented in Fig.~\ref{schem4}(c), from which the photon emission direction can be deduced, because an ultrarelativistic electron emits along the momentum direction. For a longer laser pulse the angular distribution of photon emission is broader, but, the MPER only concentrates in a narrow range near the $-z$ axis, where $\chi$ is very strong. When the electron energy approaches the condition $\gamma \approx \xi/2$ due to radiation loss, the electron could be reflected by the laser pulse [see the loop in Fig.~\ref{schem4}(d)]. Consequently, the backwards emission spectra can be observed even though its intensity is rather small compared with that in the MPER, and it also exits in the classical RR regime with a much narrower parameter range \cite{DiPiazza_2009}. The insets in Fig.~\ref{schem4}(c) and (d) clearly show that the polar angle corresponding to the $\chi$-maximum is largest for the case of $\tau_0=1.5T_0$, which is consistent with the $\theta_b$-behaviour of the MPER. When the electron moves towards the laser pulse center, the $\xi$-parameter 
keeps increasing from zero to its peak, while the electron $\gamma$-factor continuously reduces due to the radiation loss. Consequently, the parameter $\chi \propto \xi \gamma$ achieves the maximum at an intermediate laser phase $\eta_m$ prior to the laser pulse peak, and, the electron transverse momentum at this moment is $p_{\perp m}\sim m\xi (\eta_m)$. In a longer laser pulse the electron energy decrease is larger and, consequently, the $\chi$-maximum is achieved during a longer way to the peak of the pulse, i.e., $\xi (\eta_m)$ and $p_{\perp m}\sim m\xi (\eta_m)$ are smaller, yielding a reduction of the emission angle $\theta\sim p_{\perp m}/p_{\parallel m}$.
This can be seen by comparing the two cases of $\tau_0=1.5T_0$ and $\tau_0=5T_0$. However, in ultrashort pulses with duration $\tau_0\lesssim 1.5T_0$, the variation gradient of the $\xi$-parameter is more significant than that of $\gamma$ due to the strong focusing effect, and the former is larger in a shorter laser pulse. Thus, 
the laser field at the $\chi$-maximum is larger for a shorter laser pulse. This is evident in comparing the emission boundary angle and the transverse distance at the $\chi$-maximum for $\tau_0=T_0$ and $\tau_0=1.5T_0$ in Fig.~\ref{schem4}(c) and (d). Moreover, the boundary angle $\theta_b$ and frequency $\omega_b$ can be analytically estimated via the polar angle of the electron momentum and the cut-off frequency $\omega_c$, respectively, at the moment when the analytical radiation loss rate d$\varepsilon$/d$\eta$ is largest (i.e., $\chi$ is maximal), as shown in Fig.~\ref{schem2} (black dash-dotted curves with triangle marks), which qualitatively agree with the numerical results.

We proceed discussing the signatures of quantum RR in the case of an electron bunch. Following parameters are used: an electron bunch of cylindrical shape orienting along the $z$ axis, 
with a bunch length $l_e = 6\,\mu$m and a bunch radius $w_e = 3\,\mu$m, contains $N_e=10^7$ electrons. The angular spread of the bunch should be much smaller than $\Delta\theta\sim10^{\circ}$. The density of the bunch is $n_e=5.9\times 10^{16}$ cm$^{-3}$, and the relative loss of the laser energy due to Compton scattering for this electron number is estimated to be $10^{-6}$ which justifies the external field approximation for the laser field. An electron beam of such density and an energy of 500 MeV is achievable via laser-plasma acceleration in an all-optical setup \cite{Esarey_2009}. The space-charge force $F_C\sim 2\pi\alpha n_e w_e$  will be negligible with respect to the laser force $F_L\sim \xi m\omega_0$ in this case, as $F_C/F_L\sim 10^{-8}$. 
The dependences of $\theta_b$ and $\omega_b$ on the laser pulse duration for the emission of the electron bunch 
(blue solid curves with cross marks) in Fig.~\ref{schem2} remain qualitatively unaltered comparing with those of the single electron case. The RR signatures under consideration persist when the laser waist radius $w_0$ and the electron beam radius $w_e$ are within the limits of $w_0 \gtrsim 4 \lambda_0$ and $w_e \lesssim w_0/2$. Moreover, we estimate the number of laser shots to collect sufficient statistics for the observation of quantum RR signatures. 
For instance, the total probability of photon emission in the case of $\tau=5T_0$ is $W^{tot}_{ph}\approx 39.42$,
while the probability for the photon emission in d$\varepsilon$/d$\Omega|_{max}$ and d$\varepsilon$/d$\Omega|_{\varphi=\varphi_m, \theta=\theta_b}$ yields $W^{m}_{ph}\approx 0.0027$ and $W^{b}_{ph}\approx 0.007$, respectively. 
The relative signal $|W^{b}_{ph}-W^{m}_{ph}|/W^{tot}_{ph}\approx 10^{-4}$ will be larger than the statistical error $\delta_s=(W^{tot}_{ph}N_eN_{shot})^{-1/2}\sim 10^{-5}$ when the number of laser shots
is $N_{shot}=10$. The probability of a photon decaying into an electron-positron pair for the maximal $\chi_{photon}\approx0.33$ is estimated to be $10^{-4}$ and negligible.

Concluding, we have identified signatures of quantum RDR in dependence of both the angular spread and the spectral bandwidth of Compton radiation spectra on the laser pulse duration, which are distinct from those in the classical RR regime. Due to an interplay between laser beam focusing  and quantum RR effects the angular spread of the main photon emission region has a prominent maximum at an intermediate pulse duration and decreases along the further increase of the pulse duration, and, the spectral bandwidth monotonously decreases with rising pulse duration. These signatures are robust and observable in a broad range of electron and laser beam parameters.

\bibliography{strong_fields_bibliography}

\end{document}